\documentclass[10pt,a4paper]{article}

\usepackage[]{hyperref}
\usepackage{amsthm, amssymb}
\usepackage{times,mathptmx}

\def\ket#1{\left|#1\right>}

\def\expt(#1#2){\langle #2 \vert #1 \vert #2 \rangle}

\title{A discrete model of energy-conserved wavefunction collapse}

\author{Shan Gao\thanks{Institute for the History of Natural Sciences, Chinese Academy of Sciences, Beijing 100190, P. R. China. E-mail:  \href{mailto:gaoshan@ihns.ac.cn}{gaoshan@ihns.ac.cn}.}}

\begin{document}
\maketitle

\begin{abstract}\noindent

Energy nonconservation is a serious problem of dynamical collapse theories. In this paper, we propose a discrete model of energy-conserved wavefunction collapse. It is shown that the model is consistent with existing experiments and our macroscopic experience.

\end{abstract}


\vspace{6mm}

\section{Introduction}

In standard quantum mechanics, it is postulated that when the wave function of a quantum system is measured by a macroscopic device, it no longer follows the linear Schr\"{o}dinger equation, but instantaneously and randomly collapses to one of the wave functions that correspond to definite measurement results. However, this collapse postulate is ad hoc, and the theory does not tell us why and how a definite measurement result emerges (Bell 1990). A promising solution to this measurement problem is dynamical collapse theories, in which the collapse evolution is dynamical and integrated with the normal Schr\"{o}dinger evolution into a unified dynamics (Ghirardi 2011). However, the existing dynamical collapse models are plagued by the serious problem of energy nonconservation (Pearle 2000, 2007, 2009)\footnote{It is worth noting that there might also exist a possibility that the principle of conservation of energy is not universal and indeed violated by wavefunction collapse. One hint is that the usual proof that spacetime translation invariance leads to the conservation of energy and momentum relies on the linearity of quantum dynamics, and it does not apply to nonlinear quantum dynamics such as wavefunction collapse (Gao 2011, ch.3). We will not consider this possibility in this paper.}. For instance, in the CSL (Continuous Spontaneous Localization) model, the collapse due to an external noise field narrows the wave function in position space, thereby producing an increase of energy. Although it is expected that the conservation laws may be satisfied when the contributions of the noise field to the conserved quantities are taken into account (Pearle 2004; Bassi, Ippoliti and Vacchini 2005), a complete solution has not yet been found, and it is still unknown whether such a solution indeed exists. 

In this paper, we will propose a discrete model of energy-conserved wavefunction collapse. It has been demonstrated that the energy-driven collapse models that conserve energy cannot explain the emergence of definite measurement results (Pearle 2004). However, this important result does not imply that all energy-conserved collapse models are inconsistent with experiments. A detailed analysis of this paper will show that the suggested energy-conserved collapse model can be consistent with existing experiments and our macroscopic experience. The key is to assume that the energy uncertainty driving the collapse of the entangled state of a many-body system is not the uncertainty of the total energy of all sub-systems as the energy-driven collapse models assume, but the sum of the absolute energy uncertainty of every sub-system. 

\section{A discrete model of energy-conserved wavefunction collapse}

Consider a multi-level system with a constant Hamiltonian. Its initial state is:  

\begin{equation}
\ket{\psi(0)}=\sum_{i=1}^m{c_i(0) \ket{E_i}},
\label{}
\end{equation}

\noindent  where $\ket{E_i}$ is the energy eigenstate of the Hamiltonian of the system, $E_i$ is the corresponding energy eigenvalue, and $c_i(0)$ satisfies the normalization relation $\sum_{i=1}^m{|c_i(0)|^2}=1$. 

It is assumed that this superposition of energy eigenstates collapses to one of the eigenstates after a dynamical process, and the collapse evolution satisfies the conservation of energy at the ensemble level\footnote{It can be proved that only when the preferred bases (i.e. the states toward which the collapse tends) are energy eigenstates of the total Hamiltonian for each identical system in an ensemble, can energy be conserved at the ensemble level for wavefunction collapse (see Pearle 2000 for a more detailed analysis). Note that for the linear Schr\"{o}dinger evolution under an external potential, energy is conserved but momentum is not conserved even at the ensemble level, and thus it is not conservation of momentum but conservation of energy that is a more universal restriction for wavefunction collapse. Besides, as we will show later, existing experiments and our macroscopic experience only require that a superposition of energy eigenstates collapses to one of them when they are sufficiently separated in space.}. Moreover, this collapse process is composed of discrete tiny collapses\footnote{According to Gao (2011), the wave function can be regarded as a representation of the state of random discontinuous motion of particles, and the tiny collapses of the wave function may originate from the random motion of particles.}. The properties of the tiny collapses are assumed as follows. At each discrete instant $t=n t_P$ (where $t_P$ is the discrete unit of time), the probability of the tiny collapse happening in each energy branch $\ket{E_{i}}$ is equal to $P_i(t) \equiv |c_i(t)|^2$, and this collapse slightly increases the probability of the energy branch and decreases the probabilities of all other branches pro rata\footnote{It has been recently argued that the instability of the Schr\"{o}dinger evolution in the presence of a tiny perturbation of the external potential may result in the collapse of the wave function in some cases (Landsman and Reuvers 2012). It will be interesting to see whether there exists a possible connection between these tiny collapses and the perturbation.}. Then during a time interval much larger than $t_P$, the probability of each energy branch will undergo a discrete and stochastic evolution. In the end, as we will show below, the probability of one branch will be close to one, and the probabilities of other branches will be close to zero. In other words, the initial superposition will randomly collapse to one of the energy branches in the superposition. Since it has been widely conjectured that the Planck scale is the minimum spacetime scale (see, e.g. Garay 1995  for a review), we will assume that the size of each discrete instant, $t_P$, is the Planck time in our following analysis.

Now we will give a concrete analysis of this dynamical collapse process. Since the linear Schr\"{o}dinger evolution does not change energy probability distribution, we may only consider the influence of dynamical collapse on the distribution. Suppose at a discrete instant $t=n t_P$ the tiny collapse randomly happens in an energy branch $\ket{E_{i}}$, and the probability of the branch, $P_{i}(t)$, changes to

\begin{equation}
P_{i}^i (t+t_P)=P_{i}(t)+\Delta P_{i},
\label{}
\end{equation} 

\noindent where the superscript $i$ denotes this tiny collapse event, and $\Delta P_i$ is a functional of $P_{i}(t)$. Due to the conservation of probability, the probability of another branch $P_{j}(t)$ ($j \neq i$) correspondingly turns to be\footnote{One can also obtain this result by first increasing the probability of one branch and then normalizing the probabilities of all branches. This means that $P_{i}(t+t_P)={{P_{i}(t)+\Delta }\over {1+\Delta }}$ and $P_{j}(t+t_P)={{P_{j}(t)}\over {1+\Delta}}$ for any $j \neq i$. In this way, we have $\Delta P_{i}={{\Delta}\over{1+\Delta}}(1-P_{i}(t))$ and $\Delta P_{j}={{\Delta}\over{1+\Delta}}P_{j}(t)$ for any $j \neq i$. }

\begin{equation}
P_{j}^i (t+t_P)=P_{j}(t)-{{P_{j}(t) \Delta P_{i}} \over {1-P_{i}(t)}},
\label{}
\end{equation} 

\noindent where the superscript $i$ still denotes the random event. The probability of this tiny collapse happening in the energy branch $\ket{E_{i}}$ at the instant is $p(E_i,t)=P_{i}(t)$. Then we can work out the diagonal density matrix elements of the evolution\footnote{The density matrix describes the ensemble of states which arise from all possible random events.}: 

\begin{eqnarray}
\rho_{ii}(t+t_P)&=&\sum_{j=1}^m p(E_j,t) P_{i}^j (t+t_P)\nonumber
\\&=&P_{i}(t)[P_{i}(t)+\Delta P_{i}]+\sum_{j \neq i} P_{j}(t)[P_{i}(t)-{{P_{i}(t) \Delta P_{j}(t)} \over {1-P_{j}(t)}}]\nonumber
\\ &=&\rho_{ii}(t)+P_{i}(t)[\Delta P_{i}-\sum_{j \neq i} P_{j}(t){{\Delta P_{j}(t)} \over {1-P_{j}(t)}}].
\label{}
\end{eqnarray}

Here we shall introduce the first rule of dynamical collapse in our model, which says that the probability distribution of energy eigenvalues for an ensemble of identical systems is constant during the dynamical collapse process. It can be seen that this rule is entailed by the principle of conservation of energy at the ensemble level. By this rule, we have $\rho_{ii}(t+t_P)=\rho_{ii}(t)$ for any $i$. This leads to the following set of equations:

\begin{eqnarray}
\Delta P_{1}(t)-\sum_{j \neq 1}{{P_{j}(t)\Delta P_{j}(t)} \over {1-P_{j}(t)}}=0, \nonumber
\\
\Delta P_{2}(t)-\sum_{j \neq 2}{{P_{j}(t)\Delta P_{j}(t)} \over {1-P_{j}(t)}}=0, \nonumber
\\
...\nonumber
\\
\Delta P_{m}(t)-\sum_{j \neq m}{{P_{j}(t)\Delta P_{j}(t)} \over {1-P_{j}(t)}}=0.
\end{eqnarray}

\noindent By solving this equations set (e.g. by subtracting each other), we find the following relation for any $i$:

\begin{equation}
{{\Delta P_{i}} \over {1-P_{i}(t)}}=k,
\label{MK}
\end{equation}

\noindent where $k$ is an undetermined dimensionless quantity that relates to the state $\ket{\psi(t)}$. 

By using Eq. (\ref{MK}), we can further work out the non-diagonal density matrix elements of the evolution. But it is more convenient to calculate the following variant of non-diagonal density matrix elements:

\begin{eqnarray}
  \rho_{ij}(t+t_P)&=&\sum_{l=1}^m p(E_l,t) P_{i}^l (t+t_P)P_{j}^l (t+t_P)\nonumber
\\&=& \sum_{l \neq i,j} P_{l}(t)[P_{i}(t)-kP_{i}(t)][P_{j}(t)-kP_{j}(t)]\nonumber
\\ 
& & + P_{i}(t)[P_{i}(t)+k(1-P_{i}(t))][P_{j}(t)-kP_{j}(t)]\nonumber
\\ 
& & + P_{j}(t)[P_{j}(t)+k(1-P_{j}(t))][P_{i}(t)-kP_{i}(t)]\nonumber
\\ 
&=& (1-k^2)\rho_{ij}(t).
\end{eqnarray}

\noindent  Since the usual collapse time, $\tau_c$, is defined by the relation $\rho_{ij}(\tau_c)={1 \over 2}\rho_{ij}(0)$, we may use a proper approximation, where $k$ is assumed to be the same as its initial value during the time interval $[0, \tau_c]$, to simplify the calculation of the collapse time. Then we have:

\begin{equation}
\rho_{ij}(t)\approx (1-k^2)^{n}\rho_{ij}(0).
\label{}
\end{equation}

\noindent The corresponding collapse time is in the order of:

\begin{equation}
\tau_c \approx {1 \over {k^2}}t_P,
\label{TC}
\end{equation}

In the following, we shall analyze the formula of $k$ defined by Eq. (\ref{MK}). To begin with, the probability restricting condition $0\leqslant P_{i}(t)\leqslant 1$ for any $i$ requires that $0\leqslant k \leqslant 1$. When $k=0$, no collapse happens, and when $k=1$, collapse happens instantaneously. Note that $k$ cannot be smaller than zero, as this will lead to the negative value of $P_{i}(t)$ in some cases. For instance, when $k$ is negative and $P_{i}(t)< {{|k|}\over{1+|k|}}$, $P_{i}(t+t_P)=P_{i}(t)+k[1-P_{i}(t)]$ will be negative and violate the probability restricting condition. That $k$ is positive indicates that at each discrete instant only the probability of one branch increases and the probabilities of all other branches decrease, which is consistent with our previous assumption.

Next, it can be arguably assumed that $k$ is proportional to the duration of a discrete instant, namely $k \varpropto t_P$. According to Gao (2011), the discreteness of time may be a possible cause of the energy-conserved wavefunction collapse; when the duration of an instant is zero, no collapse happens, while when the duration of an instant is not zero, collapse happens. Thirdly, $k$ is also proportional to the energy uncertainty of the superposition of energy eigenstates. When the energy uncertainty is zero, i.e., when the state is an energy eigenstate, no collapse happens. When the energy uncertainty is not zero, collapse happens. How to define the energy uncertainty? Since $k$ is invariant under the swap of any two branches ($P_i, E_i$) and ($P_j, E_j$) according to Eq. (\ref{MK}), the most natural definition of the energy uncertainty of a superposition of energy eigenstates is\footnote{Since the common RMS (mean square root) uncertainty also satisfies the swap symmetry, it still needs to be studied what the exact form of $k$ is.}:

\begin{equation}
\Delta E ={1 \over 2}\sum_{i,j=1}^{m} {P_iP_j|E_{i}-E_{j}|}.
\label{EMUL}
\end{equation}

\noindent For the simplest two-level system, we have

\begin{equation}
\Delta E =P_1P_2|E_1-E_2|.
\label{ETWO}
\end{equation}

Then after omitting a coefficient in the order of unity, we can get the formula of $k$ in the first order:

\begin{equation}
k \approx \Delta E t_P/\hbar.
\label{K}
\end{equation}

\noindent This is the second rule of dynamical collapse in our model\footnote{Note that only one universal constant, the Planck time, is needed to specify the suggested collapse model. By contrast, two parameters, which were originally introduced by Ghirardi, Rimini and Weber (1986), are needed to specify the GRW and CSL models. They are a distance scale, $a \approx 10^{−5}$cm, characterising the distance beyond which the collapse becomes effective, and a time scale,  $\lambda^{-1} \approx 10^{16}$sec, giving the rate of collapse for a microscopic system. However, it is also worth noting that fundamentally these two parameters can be written in terms of other physical constants, and the CSL theory depends essentially only on one parameter, the product $\lambda a$ (Pearle and Squires 1996).}. It is worth pointing out that $k$ must contain the first order term of $ \Delta E$. For the second order or higher order term of $ \Delta E$ will lead to much longer collapse time for some common measurement situations, which contradicts experiments (Gao 2006). 

By inputting Eq. (\ref{K}) into Eq. (\ref{TC}), we can further get the collapse time formula:

\begin{equation}
\tau_c \approx {\hbar E_P \over {(\Delta E)^2}},
\label{CT}
\end{equation}

\noindent where $E_P=h/t_P$ is the Planck energy, and $\Delta E$ is the energy uncertainty of the initial state.

Based on the above analysis, the state of the multi-level system at instant $t=nt_P$ will be:

\begin{equation}
\ket{\psi(t)}=\sum_{i=1}^m{c_i(t) e^{-iE_it/\hbar}\ket{E_i}},
\label{}
\end{equation}

\noindent  Besides the linear Schr\"{o}dinger evolution, the collapse dynamics adds a discrete stochastic evolution for $P_i(t) \equiv |c_i(t)|^2$:

\begin{equation}
P_i(t+t_P) = P_i(t) +{\Delta E \over E_P} [\delta_{E_sE_i}-P_i(t)],
\label{EDC}
\end{equation}

\noindent  where $\Delta E$ is the energy uncertainty of the state at instant $t$ defined by Eq. (\ref {EMUL}), $E_s$ is a random variable representing the branch where the discrete tiny collapse happens, and its probability of assuming $E_i$ at instant $t$ is $P_i(t)$. When $E_s=E_i$, $\delta_{E_sE_i}=1$, and when $E_s \neq E_i$, $\delta_{E_sE_i}=0$.

This equation of dynamical collapse can be extended to the entangled states of a many-body system. The difference only lies in the definition of the energy uncertainty $\Delta E$. It is assumed that for a non-interacting or weakly-interacting many-body system in an entangled state, for which the energy uncertainty of each sub-system can be properly defined, $\Delta E$ is the sum of the absolute energy uncertainty of all sub-systems, namely

\begin{equation}
\Delta E={1 \over 2}\sum_{l=1}^n \sum_{i,j=1}^m {P_iP_j|E_{li}-E_{lj}|},
\label{E}
\end{equation}

\noindent where $n$ is the total number of the entangled sub-systems, $m$ is the total number of energy branches in the entangled state, and $E_{li}$ is the energy of sub-system $l$ in the \emph{i}-th energy branch of the state. Correspondingly, the final states of collapse are the product states of the energy eigenstates of the Hamiltonian of each sub-system. 

Here it should be stressed that $\Delta E$ is not defined as the uncertainty of the total energy of all sub-systems as in the energy-driven collapse models (see, e.g. Percival 1995, 1998; Hughston 1996). For each sub-system has its own energy uncertainty that drives its collapse, and the total driving ``force" for the whole entangled state should be the sum of the driving ``forces" of all sub-systems, at least in the first order approximation. Although these two kinds of energy uncertainty are equal in numerical values in some cases (e.g. for a strongly-interacting many-body system),  there are also some cases where they are not equal. For example, for a superposition of degenerate energy eigenstates of a non-interacting many-body system, which may arise during a common measurement process, the uncertainty of the total energy of all sub-systems is exactly zero, but the absolute energy uncertainty of each sub-system and their sum may be not zero. As a result, the superposition of degenerate energy eigenstates of a many-body system may also collapse. As we will see later, this is an important feature of our model, which can avoid Pearle's (2004) serious objections to the energy-driven collapse models.

It can be seen that the equation of dynamical collapse, Eq.(\ref{EDC}), has an interesting property, scale invariance. After one discrete instant $t_P$, the probability increase of the branch $\ket{E_i}$ is $\Delta P_i = {\Delta E \over E_P} (1- P_i)$, and the probability decrease of the neighboring branch $\ket{E_{i+1}}$ is $\Delta P_{i+1} = {\Delta E \over E_P} P_{i+1}$. Then the probability increase of these two branches is

\begin{equation}
\Delta (P_i+P_{i+1}) = {\Delta E \over E_P} [1-(P_i+P_{i+1})].
\label{}
\end{equation}

\noindent Similarly, the equation $\Delta P = {\Delta E \over E_P} (1- P)$ holds true for the total probability of arbitrarily many branches (one of which is the branch where the tiny collapse happens). This property of scale invariance may simplify the analysis in many cases. For instance, for a superposition of two wavepackets with energy difference, $\Delta E_{12}$, much larger than the energy uncertainty of each wavepacket, $\Delta E_1=\Delta E_2$ , we can calculate the collapse dynamics in two steps. First, we use Eq.(\ref{EDC}) and Eq.(\ref{ETWO}) with $|E_1-E_2|=\Delta E_{12}$ to calculate the time of the superposition collapsing into one of the two wavepackets. Here we need not to consider the almost infinitely many energy eigenstates constituting each wavepacket and their probability distribution. Next, we use Eq.(\ref{EDC}) with $\Delta E=\Delta E_1$ to calculate the time of the wavepacket collapsing into one of its energy eigenstates. In general, this collapse process is so slow that its effect can be neglected. 

Lastly, we want to stress another important point. In our model, the energy eigenvalues are assumed to be discrete for any quantum system. This result seems to contradict quantum mechanics, but when considering that our universe has a finite size (i.e. a finite event horizon), the momentum and energy eigenvalues of any quantum system in the universe may be indeed discrete. The reason is that all quantum systems in the universe are limited by the finite horizon, and thus no free quantum systems exist in the strict sense. For example, the energy of a massless particle (e.g. photon) can only assume discrete values $E_{n} = n^2 {hc \over 4 R_U}$, and the minimum energy is $E_{1} ={hc \over 4 R_U} \approx 10^{-33}eV$, where $R_U \approx 10^{25}m$ is the radius of the horizon of our universe. Besides, for a free particle with mass $m_0$, its energy also assumes discrete values $E_n=n^2 {h^2 \over 32m_0R_U^2}$. For instance, the minimum energy is $E_{1} \approx 10^{-72}eV$ for free electrons, which is much smaller than the minimum energy of photons\footnote{Whether this heuristic analysis is (approximately) valid depends on the application of the final theory of quantum gravity to our finite universe. However, it is worth noting that the existence of discrete energy levels for a free quantum system limited in our universe is also supported by the hypothetical holographic principle, which implies that the total information within a universe with a finite event horizon is finite. If the energy of a quantum system is continuous, then the information contained in the system will be infinite.}.

It is interesting to see whether this tiny discreteness of energy makes the collapse dynamics more abrupt. Suppose the energy uncertainty of a quantum state is $\Delta E \approx 1eV$, and its energy ranges between the minimum energy $E_1$ and $1eV$. Then we can get the maximum energy level $l_{max} \approx \sqrt{1eV\over {10^{-33}eV}} \approx 10^{16}$. The probability of most energy eigenstates in the superposition will be about $P \approx 10^{-16}$. During each discrete instant $t_P$, the probability increase of the energy branch with tiny collapse is $\Delta P \approx {\Delta E \over E_P}(1- P) \approx 10^{-28}$. This indicates that the probability change during each discrete instant is still very tiny. Only when the energy uncertainty is larger than $10^{23}eV$ or $10^{-5}E_P$, will the probability change during each discrete instant be sharp. Therefore, the collapse evolution is still very smooth for the quantum states with energy uncertainty much smaller than the Planck energy. 

\section{On the consistency of the model and experiments}

In this section, we will analyze whether the discrete model of energy-conserved wavefunction collapse is consistent with existing experiments and our macroscopic experience. Note that Adler (2002) has already given a detailed consistency analysis in the context of energy-driven collapse models, and as we will see below, some of his analysis also applies to our model.

\subsection{Maintenance of coherence}

First of all, the model satisfies the constraint of predicting the maintenance of coherence when this is observed. Since the energy uncertainty of the state of a microscopic particle is very small in general, its collapse will be too slow to have any detectable effect in present experiments on these particles. For example, the energy uncertainty of a photon emitted from an atom is in the order of $10^{-6}eV$, and the corresponding collapse time is $10^{25}s$ according to Eq. (\ref{CT}) of our collapse model, which is much longer than the age of the universe, $10^{17}s$. This means that the final states of collapse (i.e. energy eigenstates) are never reached for a quantum system with small energy uncertainty even during a time interval as long as the age of the universe. As another example, consider the SQUID experiment of Friedman et al (2000), where the coherent superpositions of macroscopic states consisting of oppositely circulating supercurrents are observed\footnote{Note that the possibility of using the SQUID experiments to test the collapse theories has been discussed in great detail by Rae (1990) and Buffa, Nicrosini and Rimini (1995).}. In the experiment, each circulating current corresponds to the collective motion of about $10^9$ Cooper pairs, and the energy uncertainty is about $8.6 \times 10^{-6}eV$. Eq. (\ref{CT})  predicts a collapse time of $10^{23}s$, and thus maintenance of coherence is expected despite the macroscopic structure of the state\footnote{A more interesting example is provided by certain long-lived nuclear isomers, which have large energy gaps from their ground states (see Adler 2002 and references therein). For example, the metastable isomer of $^{180}$Ta, the only nuclear isomer to exist naturally on earth, has a half-life of more than $10^{15}$ years and an energy gap of $75keV$ from the ground state. According to Eq. (\ref{CT}), a coherent superposition of the ground state and metastable isomer of $^{180}$Ta will spontaneously collapse to either the isomeric state or the ground state, with a collapse time of order 20 minutes. It will be a promising way to test our collapse model by examining the maintenance of coherence of such a superposition.}. 

\subsection{Rapid localization in measurement situations}

In the following, we will show that the discrete model of energy-conserved wavefunction collapse can account for the emergence of definite measurement results. 

Consider a typical measurement process in quantum mechanics. According to the standard von Neumann procedure, measuring an observable $A$ in a quantum state $\ket{\psi}$ involves an interaction Hamiltonian

\begin{equation}
H_I = g(t)PA
\label{H_int}
\end{equation} 

\noindent coupling the measured system to an appropriate measuring device, where $P$ is the conjugate momentum of the pointer variable. The time-dependent coupling strength $g(t)$ is a smooth function normalized to $\int dt g(t)=1$ during the interaction interval $\tau$, and $g(0)=g(\tau)=0$. The initial state of the pointer is supposed to be a Gaussian wave packet of width $w_0$ centered at initial position $0$, denoted by $|\phi(0)\rangle$.

For a standard (impulsive) measurement, the interaction $H_I$  is of very short duration and so strong that it dominates the rest of the Hamiltonian (i.e. the effect of the free Hamiltonians of the measuring device and the measured system can be neglected). Then the state of the combined system at the end of the interaction can be written as

\begin{equation}
\ket{t=\tau} = e^{-{i\over\hbar} P A } \ket{\psi}  \ket{\phi(0)}.
\end{equation}

\noindent By expanding $\ket{\psi}$  in the eigenstates of $A$, $\ket{a_i}$, we obtain
 
\begin{equation}
\ket{t=\tau} = \sum_{i} e^{-{i\over\hbar} P a_i } c_i \ket{a_i} \ket{\phi(0)},
\end{equation}

\noindent where $c_i$ are the expansion coefficients. The exponential term shifts the center of the pointer by $a_i$:

\begin{equation}
\ket{t=\tau} = \sum_{i} c_i \ket{a_i} \ket{\phi(a_i)}.
\end{equation}

\noindent  This is an entangled state, where the eigenstates of $A$ with eigenvalues $a_i$ get correlated to macroscopically distinguishable states of the measuring device in which the pointer is shifted by these values $a_i$ (but the width of the pointer wavepacket is not changed). According to the collapse postulate, this state will instantaneously and randomly collapse into one of its branches $\ket{a_i} \ket{\phi(a_i)}$. Correspondingly, the measurement will obtain a definite result, $a_i$, which is one of the eigenvalues of the measured observable.

Let's see whether the energy-conserved collapse model can explain the emergence of the definite measurement results. At first sight, the answer seems negative. As stressed by Pearle (2004), each outcome state of the measuring device in the above entangled superposition has precisely the same energy spectrum for an ideal measurement\footnote{According to Pearle (2004), when considering environmental influences, each device/environment state in the superposition also has precisely the same energy spectrum.}. Then it appears that the superposition will not collapse according to the energy-conserved collapse model\footnote{As noted before, the collapse due to the tiny energy uncertainty of the measured state can be neglected.}. However, this is not the case. The key is to realize that different eigenstates of the measured observable are generally measured in different parts of the measuring device, and they interact with different groups of atoms or molecules in these parts\footnote{In the final analysis, different results are in general represented by different positions of the pointer of the measuring device, and thus they always appear in different spatial parts of the device.}. Therefore, we should rewrite the device states explicitly as $\ket{\phi(0)} = \prod_j \ket{\varphi_j(0)}$ and $\ket{\phi(a_i)}=\ket{\varphi_i(1)}\prod_{j\neq i}\ket{\varphi_j(0)}$, where $ \ket{\varphi_j(0)}$ denotes the initial state of the device in part $j$, and $\ket{\varphi_i(1)}$ denotes the outcome state of the device in part $i$. Then we have

\begin{equation}
\sum_{i} c_i \ket{a_i} \ket{\phi(a_i)} =  \sum_{i} c_i \ket{a_i} \ket{\varphi_i(1)}\prod_{j\neq i}\ket{\varphi_j(0)}.
\end{equation}

\noindent Since there is always some kind of measurement amplification from the microscopic state to the macroscopic outcome in the measurement process, there is a large energy difference between the states $\ket{\varphi_i(1)}$ and $\ket{\varphi_i(0)}$ for any $i$.\footnote{Since each outcome state of the measuring device has the same energy spectrum, the energy difference between the states $\ket{\varphi_i(1)}$ and $\ket{\varphi_i(0)}$ is the same for any $i$.} As a result, the total energy uncertainty, which is approximately equal to the energy difference according to Eq. (\ref{E}), is also very large, and it will result in a rapid collapse of the above superposition into one of its branches according to the energy-conserved collapse model\footnote{Since the uncertainty of the total energy of the whole entangled system is still zero, the energy-driven collapse models (e.g. Percival 1995; Hughston 1996) will predict that no wavefunction collapse happens and no definite measurement result emerges for the above measurement process (Pearle 2004).}. 

Let's give a more realistic example, a photon being detected via photoelectric effect. In the beginning of the detection, the spreading spatial wave function of the photon is entangled with the states of a large number of surface atoms of the detector. In each local branch of the entangled state, the total energy of the photon is wholly absorbed by the electron in the local atom interacting with the photon\footnote{In more general measurement situations, the measured particle (e.g. an electron) is not annihilated by the detector. However, in each local branch of the entangled state of the whole system, the particle also interacts with a single atom of the detector by an ionizing process, and energy also conserves during the interaction. Due to this important property, although the measured particle is detected locally in a detector (the size of the local region is in the order of the size of an atom), its wave function does not necessarily undergo position collapse as assumed by the GRW and CSL models etc, and especially, energy can still be conserved during the localization process according to our model.}. This is clearly indicated by the term $\delta (E_f-E_i-\hbar \omega)$ in the transition rate of photoelectric effect. The state of the ejecting electron is a (spherical) wavepacket moving outward from the local atom, whose average direction and momentum distribution are determined by the momentum and polarization of the photon.

This microscopic effect of ejecting electron is then amplified (e.g. by an avalanche process of atoms) to form a macroscopic signal such as the shift of the pointer of a measuring device. During the amplification process, the energy difference is constantly increasing between the branch in which the photon is absorbed and the branch in which the photon is not absorbed near each atom interacting with the photon. This large energy difference will soon lead to the collapse of the whole superposition into one of the local branches, and thus the photon is only detected locally. Take the single photon detector - avalanche photodiode as a concrete example. Its energy consumption is sharply peaked in a very short measuring interval. One type of avalanche photodiode operates at $10^5$ cps and has a mean power dissipation of 4mW (Gao 2006). This corresponds to an energy consumption of about $2.5\times 10^{11}eV$ per measuring interval $10^{-5}s$. By using the collapse time formula Eq. (\ref{CT}), where the energy uncertainty is $\Delta E \approx 2.5\times 10^{11}eV$, we find the collapse time is  $\tau_c \approx 1.25\times 10^{-10}s$. This collapse time is much shorter than the measuring interval.

\subsection{Emergence of the classical world}

In this subsection, we will show that the discrete model of energy-conserved wavefunction collapse is also consistent with our macroscopic experience. 

At first glance, it appears that there is an apparent inconsistency. According to the model, when there is a superposition of a macroscopic object in an identical physical state (an approximate energy eigenstate) at two different, widely separated locations, the superposition does not collapse, as there is no energy difference between the two branches of the superposition. But the existence of such superpositions is obviously inconsistent with our macroscopic experience; macroscopic objects are localized. This common objection has been basically answered by Adler (2002). The crux of the matter lies in the influences of environment. The collisions and especially the accretions of environmental particles will quickly increase the energy uncertainty of the entangled state of the whole system including the object and environmental particles, and thus the initial superposition will soon collapse to one of the localized branches according to our model. Accordingly, the macroscopic objects can always be localized due to environmental influences. It should be stressed again that the energy uncertainty here denotes the sum of the absolute energy uncertainty of each sub-system in the entangled state as defined in our model\footnote{The uncertainty of the total energy of the whole system is still very small even if the influences of environment are counted. Thus no observable collapse happens for the above situation according to the energy-driven collapse models (Pearle 2004).}.

As a typical example, we consider a dust particle of radius $a \approx 10^{-5}cm$ and mass $m \approx 10^{-7}g$. It is well known that localized states of macroscopic objects spread very slowly under the free Schr\"{o}dinger evolution. For instance, for a Gaussian wave packet with initial (mean square) width $\Delta$, the wave packet will spread so that the width doubles in a time $t=2m \Delta^2 / \hbar $. This means that the double time is almost infinite for a macroscopic object. If the dust particle had no interactions with environment and its initial state is a Gaussian wave packet with width $\Delta \approx 10^{-5}cm$, the doubling time would be about the age of the universe. However, if the dust particle interacts with environment, the situation turns out to be very different. Although the different components that couple to the environment will be individually incredibly localised, collectively they can have a spread that is many orders of magnitude larger. In other words, the state of the dust particle and the environment will be a superposition of zillions of very well localised terms, each with slightly different positions, and which are collectively spread over a macroscopic distance (Bacciagaluppi 2008). According to Joos and Zeh (1985), the spread in an environment full of thermal radiation only is proportional to mass times the cube of time for large times, namely $(\Delta x) ^2 \approx \Lambda m\tau^3$, where $\Lambda$ is the localization rate depending on the environment, defined by the evolution equation of density matrix $\rho_t(x,x')=\rho_0(x,x')e^{-\Lambda t (x-x')^2}$. For example, if the above dust particle interacts with thermal radiation at $T=300K$, the localization rate is $\Lambda = 10^{12}$, and the overall spread of its state is of the order of $10m$ after a second (Joos and Zeh 1985). If the dust particle interacts with air molecules, e.g. floating in the air, the spread of its state will be much faster. 

Let's see whether the energy-conserved collapse in our model can prevent the above spreading. Suppose the dust particle is in a superposition of two identical localized states that are separated by $10^{-5}cm$ in space. The particle floats in the air, and its average velocity is about zero. At standard temperature and pressure, one nitrogen molecule accretes in the dust particle, whose area is $10^{-10}cm^2$, during a time interval of $10^{-14}s$ in average (Adler 2002). Since the mass of the dust particle is much larger than the mass of a nitrogen molecule, the change of the velocity of the particle is negligible when compared with the change of the velocity of the nitrogen molecules during the process of accretion. Then the kinetic energy difference between an accreted molecule and a freely moving molecule is about $\Delta E = {3 \over 2}kT \approx 10^{-2}eV$. When one nitrogen molecule accretes in one localized branch of the dust particle (the molecule is freely moving in the other localized branch), it will increase the energy uncertainty of the total entangled state by $\Delta E \approx 10^{-2}eV$. Then after a time interval of $10^{-4}s$, the number of accreted nitrogen molecules is about $10^{10}$, and the total energy uncertainty is about $10^{8}eV$. According to Eq. (\ref{CT}) of our collapse model, the corresponding collapse time is about $10^{-4}s$. 

In the energy-conserved collapse model, the final states of collapse are energy eigenstates, and in particular, they are nonlocal momentum eigenstates for free quantum systems. Thus it is somewhat counterintuitive that the energy-conserved collapse can make the states of macroscopic objects local. As shown above, this is due to the constant influences of environmental particles. When the spreading of the state of a macroscopic object becomes larger, its interaction with environmental particles will introduce larger energy difference between its different local branches, and this will then collapse the spreading state again into a more localized state\footnote{It is interesting to note that the state of a macroscopic object can also be localized by the linear Schr\"{o}dinger evolution via interactions with environment, e.g. by absorbing an environmental particle with certain energy uncertainty. For example, if a macroscopic object absorbs a photon (emitted from an atom) with momentum uncertainty of $\Delta p \approx 10^{-6}eV/c$, the center-of-mass state of the object, even if being a momentum eigenstate initially, will have the same momentum uncertainty by the linear Schr\"{o}dinger evolution, and thus it will become a localized wavepacket with width about $0.1m$. Note that there is no vicious circle here. The energy spreading state of a microscopic particle can be generated by an external potential (e.g. an electromagnetic potential) via the linear Schr\"{o}dinger evolution, and especially they don't necessarily depend on the localization of macroscopic objects such as measuring devices. Thus we can use the existence of these states to explain the localization of macroscopic objects.}. As a result, the states of macroscopic objects in an environment will never reach the final states of collapse, namely momentum eigenstates, though they do continuously undergo the energy-conserved collapse. To sum up, there are two opposite processes for a macroscopic object constantly interacting with environmental particles. One is the spreading process due to the linear Schr\"{o}dinger evolution, and the other is the localization process due to the energy-conserved collapse evolution. The interactions with environmental particles not only make the spreading more rapidly but also make the localization more frequently. In the end these two processes will reach an approximate equilibrium. The state of a macroscopic object will be a wave packet narrow in both position and momentum, and this narrow wave packet will approximately follow Newtonian trajectories by Ehrenfest's theorem (if the external potential is uniform enough along the width of the packet)\footnote{When assuming the energy uncertainty of an object is in the same order of its thermal energy fluctuation, we can estimate the rough size of its wavepacket. For instance, for a dust particle of mass $m=10^{-7}g$, its root mean square energy fluctuation is about $10^3eV$ at room temperature $T=300K$ (Adler 2002), and thus the width of its wavepacket is about $10^{-10}m$.}. In some sense, the emergence of the classical world around us is ``conspired" by environmental particles according to the energy-conserved collapse model.

\subsection{Definiteness of our conscious experiences}

Ultimately, the energy-conserved collapse model should be able to account for our definite conscious experiences. According to recent neuroscience literature, the appearance of a (definite) conscious perception in human brains involves a large number of neurons changing their states from resting state (resting potential) to firing state (action potential). In each neuron, the main difference of these two states lies in the motion of $10^6$ $Na^+$s passing through the neuron membrane. Since the membrane potential is in the order of $10^{-2}V$, the energy difference between firing state and resting state is $\Delta E \approx 10^{4}eV$. According to Eq. (\ref{CT}) of the energy-conserved collapse model, the collapse time of a quantum superposition of these two states of a neuron is $\tau_c \approx 10^{5}$s. When considering the number of neurons that can form a definite conscious perception is usually in the order of $10^7$, the collapse time of the quantum superposition of two different conscious perceptions is $\tau_c  \approx 10^{-9}$s. Since the normal conscious time of a human being is in the order of several hundred milliseconds, the collapse time is much shorter than the normal conscious time. Therefore, our conscious perceptions are always definite according to the energy-conserved collapse model\footnote{Note that a serious analysis of human perceptions such as visual perception in terms of the collapse theories such as the GRW model was first given by Aicardi et al (1991) and Ghirardi (1999).}.

\section{Conclusions}

In this paper, we propose a discrete model of energy-conserved wavefunction collapse, and show that the model is consistent with existing experiments and our macroscopic experience. This provides a possible new solution to the problem of energy nonconservation for dynamical collapse theories. 

\section*{Acknowledgments}
I am very grateful to two anonymous reviewers for their insightful comments, constructive criticisms and helpful suggestions.

\section*{References}
\renewcommand{\theenumi}{\arabic{enumi}}
\renewcommand{\labelenumi}{[\theenumi]}
\begin{enumerate}

\item Adler, S. L. (2002). Environmental influence on the measurement process in stochastic reduction models, J. Phys. A: Math. Gen. 35, 841-858.
\item Aicardi, F., Borsellino, A., Ghirardi, G. C., and Grassi, R. (1991). Dynamical models for state-vector reduction: Do they ensure that measurements have outcomes? Found. Phys. Lett. 4, 109.
\item Bacciagaluppi, G. (2008). The role of decoherence in quantum mechanics. The Stanford Encyclopedia of Philosophy (Fall 2008 Edition), Edward N. Zalta (eds.), http://plato.stanford.edu/archives/fall2008/entries/qm-decoherence/. 
\item Bassi, A., Ippoliti, E., and  Vacchini, B. (2005). On the energy increase in space-collapse models. J. Phys. A : Math. Gen. 38, 8017.
\item Bell, J. S. (1990). Against `measurement', in A. I. Miller (eds.), Sixty-Two Years of Uncertainty: Historical Philosophical and Physics Enquiries into the Foundations of Quantum Mechanics. Berlin: Springer, 17-33.
\item Buffa, M., Nicrosini, O., and Rimini, A. (1995). Dissipation and reduction effects of spontaneous localization on superconducting states. Found. Phys. Lett. 8, 105-125.
\item Friedman, J. R., Patel, V., Chen, W., Tolpygo, S. K. and Lukens, J. E. (2000). Quantum Superposition of Distinct Macroscopic States. Nature 406, 43.
\item Gao, S. (2006). A model of wavefunction collapse in discrete space-time. Int. J. Theor. Phys. 45, 1943-1957.
\item Gao, S. (2010). On Di\'{o}si-Penrose criterion of gravity-induced quantum collapse. Int. J. Theor. Phys. 49, 849-853.
\item Gao, S. (2011). Interpreting quantum mechanics in terms of random discontinuous motion of particles. http://philsci-archive.pitt.edu/9057.
\item Garay, L. J.  (1995). Quantum gravity and minimum length. Int. J. Mod. Phys. A 10, 145.
\item Ghirardi, G. C. (1999). Quantum superpositions and definite perceptions: Envisaging new feasible experimental tests. Phys. Lett. A 262, 1-14.
\item  Ghirardi, G. C. (2011). Collapse Theories. The Stanford Encyclopedia of Philosophy (Winter 2011 Edition), Edward N. Zalta (ed.), http://plato.sta-nford.edu/archives/win2011/entries/qm-collapse/.
\item Ghirardi, G. C., Rimini, A., and Weber, T. (1986). Unified dynamics for microscopic and macroscopic systems. Phys. Rev. D 34, 470.
\item Hughston, L. P. (1996). Geometry of stochastic state vector reduction. Proc. Roy. Soc. A 452, 953.
\item Joos, E. and Zeh, H. D. (1985). The emergence of classical properties through interaction with the environment. Zeitschrift f\"{u}r Physik B 59, 223-243. 
\item Landsman, N. P. and Reuvers, R. (2012). A Flea on Schr\"{o}dinger's Cat. arXiv:1210.2353 [quant-ph].
\item Pearle, P. (2000). Wavefunction Collapse and Conservation Laws. Found. Phys. 30, 1145-1160.
\item Pearle, P. (2004). Problems and aspects of energy-driven wavefunction collapse models. Phys. Rev. A 69, 42106. 
\item Pearle, P. (2007). How stands collapse I. J. Phys. A: Math. Theor., 40, 3189-3204.
\item Pearle, P. (2009). How stands collapse II. in Myrvold, W. C. and Christian, J.  eds., Quantum Reality, Relativistic Causality, and Closing the Epistemic Circle: Essays in Honour of Abner Shimony. The University of Western Ontario Series in Philosophy of Science, 73(IV), 257-292.
\item Pearle, P. and Squires, E. (1996). Gravity, energy conservation and parameter values in collapse models. Found. Phys. 26, 291.
\item Percival, I. C. (1995). Quantum space-time fluctuations and primary state diffusion. Proc. Roy. Soc. A 451, 503.
\item Percival, I. C. (1998). Quantum State Diffusion. Cambridge: Cambridge University Press.
\item Rae, A. I. M. (1990). Can GRW theory be tested by experiments on SQUIDS? J. Phys. A: Math. Gen. 23, L57.

\end{enumerate}
\end{document}